\documentclass[journal]{IEEEtran}
\ifCLASSINFOpdf
  \usepackage[pdftex]{graphicx}
\else
\fi
\usepackage{url}
\usepackage[table,xcdraw]{xcolor}
\usepackage[justification=centering]{caption}

\begin{document}
\title{Monitor++?: Multiple versus Single Laboratory Monitors in Early Programming Education}

\author{Matthew~Stephan\\Department of Computer Science and Software Engineering, Miami University, Oxford, OH, USA, 45056}% <-this % stops a space
%\thanks{Department of Computer Science and Software Engineering, Miami University, Oxford, OH, USA, 45056 Contact:stephamd@miamioh.edu}}


\markboth{Monitor++? Multiple versus Single Laboratory Monitors in Early Programming Education}%
{Matthew Stephan}



\IEEEaftertitletext{\vspace{-2.5\baselineskip}}
\maketitle

\begin{abstract}
CONTRIBUTION: This paper presents an empirical study of an introductory-level programming course with students using multiple monitors and compares their performance and self-reported experiences versus students using a single monitor. 
BACKGROUND: Professional-level programming in many technological fields often employs multiple-monitors stations, however, some education laboratories employ single-monitor stations. This is unrepresentative of what students will encounter in practice and experiential learning.
RESEARCH QUESTIONS: This study aims to answer three research questions. The questions include discovering the experiential observations of the students, contrasting the performance of the students using one monitor versus those using two monitors, and an investigation of the ways in which multiple monitors were employed by the students.   
METHODOLOGY: Half of the students in the study had access to multiple monitors.  This was the only difference between the two study groups. This study contrasts grade medians and conducts median-test evaluation. Additionally, an experience survey facilitated likert-scale values and open-ended feedback questions facilitated textual analysis. Limitations of the study include the small sample size (86 students) and lack of control of participant composition.
FINDINGS: Students reacted very favorably in rating their experience using the intervention. Overall, the multiple-monitor group had a slight performance improvement. Most improvement was in software-design and graphics assignments. Performance increased statistically significantly on the interfaces-and-hierarchies labs.  Students used multiple-monitors in different ways including reference guides, assignment specifications, and more.\end{abstract}

%\textbf{Objective:} We ascertain student experience using multiple monitors and compare students’ performance versus single monitors. We hypothesize multiple-monitors students will react positively, while performing better than their single-monitor peers.   
%\newline
% 
%\newline
%\textbf{Implications:} Students overwhelmingly enjoyed using multiple-monitor stations. Research is needed to conclude what specific types of assignments will see improved performance when employing multiple-monitors stations. Our students performed better on design and graphics assignments, indicating a good starting point for researchers.


% \fi
\IEEEpeerreviewmaketitle

\vspace{-0.25cm}
\section{Introduction}
Software developers often build software by referring to and using community based solutions~\cite{vasilescu2013stackoverflow}.  They also increasingly collaborate with both co-located and distributed teammates~\cite{highsmith2013adaptive}. While programming, developers frequently refer to system requirements and design models to ensure their implementations adhere to specifications, especially in formal software engineering. As such, the need for additional screen real estate is paramount to effective and efficient programming. 

One approach to improve efficiency and digital visual space in many domains is employing multiple monitors in a single workspace. Studies show that using more than one monitor improves productivity. For example, researchers at the University of Utah found that students who used two 20-inch monitors were 44\% faster performing tasks, such as (word) document editing, than those with one monitor~\cite{colvin2004productivity}.  Other research demonstrates that using multiple monitors increases memory cognition~\cite{tan2001infocockpit}, and improves users' subjective experiences~\cite{grudin2001partitioning}.  Additionally, Gallagher et al. performed a literature review of studies comparing multiple to single monitor work stations focusing on productivity for work tasks and health outcomes~\cite{doi:10.1177/1071181319631210}. They found that multiple monitor stations "generally improved or remained the same" in terms of productivity compared to single monitor work stations. 

While a growing trend in open software development workspaces is to employ two or more monitors, many educational environments, such as university computer labs, have students constrained to a single monitor. This is unrepresentative of what they will be using in the real world and hinders their ability to effectively think and problem solve programmatically. For example, they are less able to perform quick context switches without disrupting the flow of each programming task.  With two or more monitors, one monitor can be used for building software components while the other(s) can display reference materials, such as programming language documentation or to display code from related software components or examples.  

Through their institution, Miami University, the author received a technology grant for improving student learning, which was used towards installing multiple monitor work stations in the main computer laboratory\footnote{\url{https://tinyurl.com/miamiMultMontiors}}. The intent was students would perform better, learn more effectively, and improve their laboratory experience. Additionally, students would now be able to learn software development in an authentic environment that closely mirrors the kind of spaces found in contemporary software development organizations. Based on the grant's funds, only the most used and largest computer laboratory could be upgraded, leaving the others with only single monitor work stations.  Thus, some sections of the courses were still constrained to single monitor learning environments. However, it did present the unique opportunity to observe using multiple monitors as an educational intervention, while having an organic control group in the form of single monitor work stations. 

This paper presents an empirical study and lab-based investigation in using multiple laboratory monitors as an instructional technology in an introductory programming course. The following specific research questions are explored in this context,

\begin{enumerate}
    \item What are the (subjective) experiences of students when using multiple monitor work stations to learn programming? 
    \item  How does the academic laboratory performance (grades) compare in students using single monitor work stations to those using multiple monitors work stations? 
    \item What are some of the ways in which students use multiple monitor work stations and in what ways do students perceive that such work stations either helped or hindered their learning? 
\end{enumerate}

This study focuses on quantitative student performance, and both qualitative and quantitative self-reported laboratory experience.  Two sections of the course used multiple monitors during their programming laboratory sessions, and two sections of the course were the control group, having access to one monitor only.  This paper presents quantitative data findings on the two groups' laboratory performance over the semester, and the subjective quantitative and qualitative feedback and impressions of the students who had access to two monitor work stations. This study will help further the case for providing all students with multiple monitor work stations and helps guide future research in this area. 

This paper begins by presenting background and related work in Sections~\ref{sec:background} and \ref{sec:related}, respectively. Section~\ref{sec:contextAnDIntervention} describes both the context of the intervention, that is, the course and students, and the multiple monitor intervention. The results of the empirical study are in Section~\ref{sec:experienceReport} including a presentation of quantitative data contrasting laboratory performance scores,  a quantitative presentation of the students' laboratory experience through a Likert-type survey, and a qualitative summary of the open-ended responses provided by the students about their experience using multiple monitors.  Analysis and insights into the data, and potential threats to validity, are in Section~\ref{sec:discussion}.  Section~\ref{sec:conlusion} concludes the paper. 
\vspace{-2mm}
\section{Background}
\label{sec:background}

While adding more visual real estate to improve efficiency appears intuitive, there is research in the literature that supports the application of this as an pedagogical instructional technology intervention. Gallager et al. performed a systematic literature review of studies that focus on the health and performance of using multiple monitors for "office work tasks" (not programming)~\cite{doi:10.1177/1071181319631210}. They found that 1) performance generally either improved or stayed the same with multiple monitors work stations, and that 2) "Health-related outcomes were less consistent". 
Tan et al. provided multiple monitors to users to test human memory in a spatial system~\cite{tan2001infocockpit}. Their user study demonstrated a 56\% increase in memory compared to single-monitor control users. The notions of location and recall are important considerations for programming education. 

Grudin performed some foundational work considering how people use space, and arrange information digitally~\cite{grudin2001partitioning}. They note that a second monitor is key for secondary activities and resources, which they call "peripheral awareness".  Users in their study were in strong favour of using multiple monitors after the experience.  Similarly, Colvin et al. performed experiments with single and multiple monitor configurations for office tasks such as working with spreadsheets, and editing slide shows and textual documents~\cite{colvin2004productivity}. They found significant efficiency improvements for the multiple monitor cases, recommending them as a cost effective measure and "for use in any situation where multiple screens of information are ordinary part of the work". The research in this paper contends that programming, and its education, falls within that description.

\section{Related Work}
\label{sec:related}

In this section, related work is aimed at deploying and evaluating multiple monitors in various settings related or analogous to programming. 

Bi and Balakrishnan conducted a week-long case study investigating "daily" office tasks such as web browsing, word processing, and emailing by typical users using a single monitor and dual monitors. The researchers evaluated the users' self-reported enjoyment and efficiency in using an ultra-large display instead~\cite{bi2009comparing}. The single- and dual- monitor subjects preferred the large display for many of the tasks.  While they do mention programming as an application, their large display is 16 feet by 6 feet, which is not practical in a university laboratory setting. 

Hutchings et al. performed research into how people interact with interfaces in multiple monitors, and how to best interact with components across different displays~\cite{hutchings2004display}.  Their research includes novel uses of multiple monitors, developing new environments for multiple monitors, and more~\cite{hutchings2004exploring}.   While none of their work focuses on education, some of their detailed interaction analysis and novel ideas are interesting areas of future work for multiple monitor programming laboratory education.

The work most related to this paper is that of Chang et al., who investigated using two different projected screens during their lectures in a programming course~\cite{chang2011comparison}.  This differs from the work in this study as they focus on lecture slides and course material presentation, and its impact on physiological aspects like cognitive load and split attention. Similarly, Renambot and Schaaf present a user interface capable of implementing tiled displays for educational presentations and lectures~\cite{renambot2002enabling}. 

\section{Materials and Methods}
\label{sec:contextAnDIntervention}

This section includes the materials and methods used in the study, including the intervention and the specific context in which it was applied. 
\vspace{-2.5mm}
\subsection{Intervention and Materials}

This pedagogical intervention involved upgrading an existing and the most used computer laboratory. The specific materials for the intervention were  outfitting 36 work stations with two 19 inch Dell monitors\footnote{\url{http://www.dell.com/ed/business/p/dell-p1917s-monitor/pd}} each.  The monitors were placed directly together and adjacent using desk mounts and a docking station.   As an intervention, it was considered how this would impact the students' laboratory experience and learning in the specific context described in the next subsection. 

The goal was to create an environment for software engineering and computer science majors that replicates, as closely as possible, a workspace that is used by practicing software developers. Due to the constraints of the grant, there were no upgrades to any other computer laboratories at that time, leaving some sections of courses constrained to a single monitor work station setup. Thus, the intervention in this context was applied to only a portion of the students, while the others belonged to the control group, organically. 

\subsection{Context and Methods}
%Spring 2017
For this instructional technological intervention was applied to an introductory object-oriented programming course, which consisted of four sections of students.  This course uses the Java programming language. Due to scheduling and demand, many sections were at the same time. As a result, two sections were able to use multiple monitor work stations for their laboratory sessions, and two were limited to the existing, pre-grant, single monitor work stations. 

The makeup of the students consisted primarily of computer science and software engineering majors. Importantly, all students were required to have a grade of C- or above in a perquisite and the preliminary (1 of 2) introduction to programming course. Thus, the students were relatively equally prepared for the course. 87 students were enrolled in the course, however, in this empirical study only 86 students consented to the study parameters. As shown in Table~\ref{tbl:summStudents}, 46 of the students had access to multiple monitors during their laboratory assignments, and 40 students did not.  For students who had access to multiple monitor work stations, they were not instructed in how to best use the monitors in this specific course offering.  The author contends that providing students instruction or training to make optimal use of such setups would confound any findings and be an unfair and unrealistic comparison.  From herein, the 46 students in the two sections with multiple monitors are represented as Group MM, and the 40 students with single monitor work stations are represented as Group SM. 

\begin{table}
\centering
\caption{Summary of Students}
\label{tbl:summStudents}
\begin{tabular}{|c|c|c|}
\hline
\textbf{Group Name} & \textbf{Group Description} & \textbf{\# Students} \\ \hline
Group MM &Multiple monitor work stations& 46 \\ \hline
Group SM &Single monitor work stations& 40 \\ \hline
Total & All students  & 86 \\ \hline
\end{tabular}
\vspace{-5mm}
\end{table}

This course offering consisted of 12 laboratory assignments throughout the semester intended to reinforce the course material, facilitate experiential learning, and assess the students' knowledge. The assignments consisted of one large or two smaller problems pertaining to the course material.  The students were given the laboratory time period to complete the assignments, and in some cases but equally for all sections, longer.  The 12 laboratory sessions were monitored by a rotation of 4 teaching assistants.  Thus, all 4 course sections had each teaching assistant 3 times throughout the semester, with each teaching assistant rotating laboratory grading synchronously. Grading and evaluation was performed by the teaching assistants against an instructor-provided rubric and overseen by the instructor. For analysis, this study performs both direct comparison of student median grades and Mood's median-test chi-squared analysis~\cite{corder2014nonparametric} on the resulting data.  

The students were presented Institutional Review Board (IRB)-approved consent forms to use the aggregate data of their laboratory grades, which they could decline at no risk. Only one student declined, and their data is not included in the data in this paper in any form as they did not complete the course regardless.   At the end of the semester, the Group MM students were asked to complete a fully optional, anonymous, and IRB-approved survey describing their subjective laboratory experiences using the intervention, using both Likert-type scale questions, and open ended qualitative questions.
\vspace{-2.5mm}
\section{Results}
\label{sec:experienceReport}

The following section presents and summarizes the data from observations.  The raw data is available online so others can learn from the intervention and the students' experiences, and use it to help with their adoption of multiple monitors in their educational laboratories\footnote{\url{http://hdl.handle.net/2374.MIA/6612}}. This section is organized and divided into the 3 types of analysis performed: quantitative laboratory performance, quantitative laboratory experience, and qualitative laboratory experience, which address research questions 1, 2, and 3, respectively.  In the subsequent section, data analysis addresses the research questions and provides some discussion points. 

%An outcome that will be immediately measurable will be to explicitly teach students how to best problem solve and implement solutions using multiple monitors.  To assess this outcome and its value, we can have them demonstrate their ability in the lab itself to harness the power of two monitors.  Seeing as much lab work must be started and finished within a single lab period, such as two hours, assessing speed/efficiency is an outcome that can be quantified.   In addition, for courses with the same instructor that use the same or similar labs from year to year, a correlation can be established in terms of quality of the software developed by the students that used one monitor in past offerings, and those that use two.
\vspace{-2.5mm}
\subsection{Quantitative - Laboratory Experience via Likert-type Scale}

The majority of the MM Group students replied to the optional survey. All of them started it, but not all of them answered every question, as they were allowed to stop any time. This section includes the resulting percentages, rounded to the nearest whole number.  Before any Likert-type scale questions, the students were first asked,
\begin{enumerate}
    \item Was this your first time using multiple monitors?
    \item Was this your first time using multiple monitors for programming?
\end{enumerate}

\begin{figure}
    \centering
    \includegraphics[width=\columnwidth]{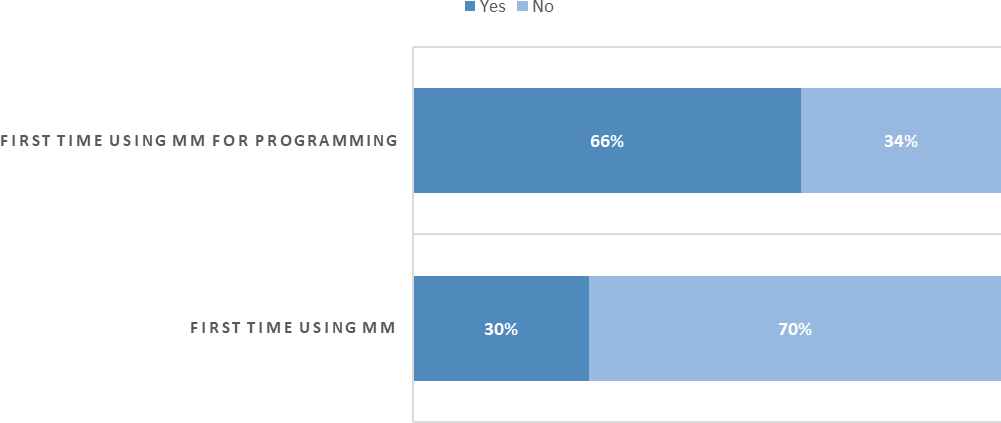}
    \caption{Survey Responses to "First Time" Questions}
    \label{fig:firstTime}
    \vspace{-5mm}
\end{figure}

As shown in Figure~\ref{fig:firstTime}, many respondents had used multiple monitors before in general, 70\%. However, an almost equally large percentage, 66\%, had \emph{not} used them for programming before. 

While not perfect, Likert-type scale questions do facilitate providing quantitative value to qualitative data~\cite{allen2007likert}. While the verdict is still out on if switching the positive or negative tone of the questions reduces acquiescence bias~\cite{friborg2006likert,johns2010likert}, this survey did so.  The survey asked questions using a 7-point scale, as recommended by Human-Computer Interaction research~\cite{lewis1993multipoint}. The statements the students were to consider included,

\begin{enumerate}
    \setcounter{enumi}{2}
    \item I would choose to use multiple monitors for programming again in the future
    \item Using multiple monitors was very difficult for me and caused me discomfort
    \item I think programming should be taught by having students use multiple monitors
    \item Multiple monitors hindered (hurt) my ability to learn this course's content
    \item I had trouble deciding which content to put on which monitor
    \item I found using multiple monitors was intuitive and easy to get used to
    \item I would not recommend using multiple monitors for programming to a colleague/peer
    \item Multiple monitors facilitated my learning, making the labs much easier to complete
\end{enumerate}

\begin{figure}
    \centering
    \includegraphics[width=\columnwidth]{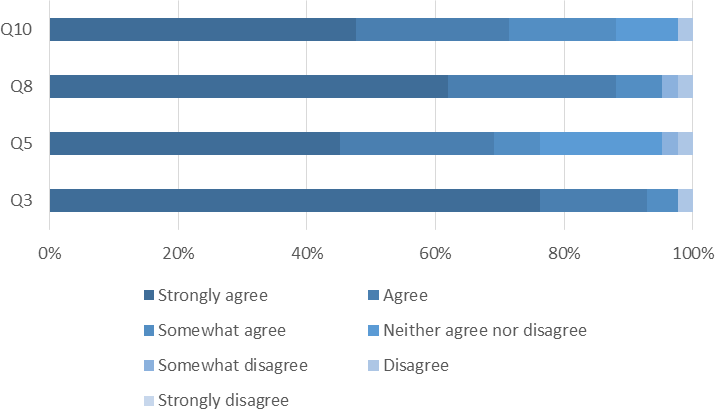}
    \caption{Responses to Positive Questions}
    \label{fig:positiveLikert}
\end{figure}

\begin{figure}
    \centering
    \includegraphics[width=\columnwidth]{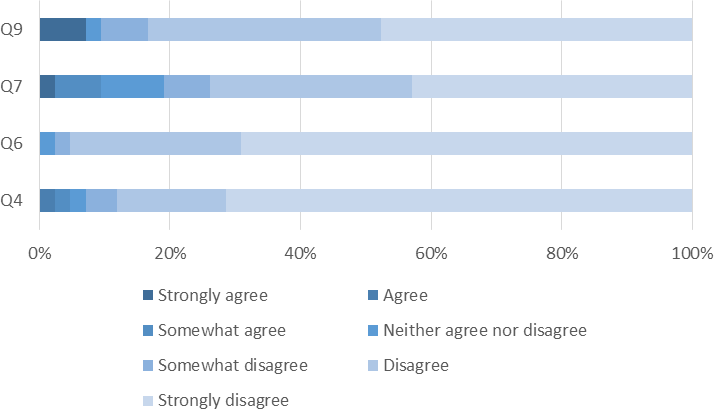}
    \caption{Responses to Negative Questions}
    \label{fig:negativeLikert}
    \vspace{-5mm}
\end{figure}

The responses to the positively phrased questions, 3; 5; 8; and 10, are summarized in Figure~\ref{fig:positiveLikert}, and the negatively phrased questions, 4; 6; 7; and 9, are summarized in Figure~\ref{fig:negativeLikert}. It is in a monochromatic scheme for accessibility, with the darker shades representing stronger agreement, and the lighter shades representing stronger disagreement.  Thus, in Figure~\ref{fig:positiveLikert} darker shades are more supportive of positive experiences in using multiple monitors.  In contrast, in Figure~\ref{fig:negativeLikert}, the lighter shades are more supportive of positive experiences using multiple monitors. Raw data is in Table~\ref{tbl:rawLikertData}. Insights on, and analysis of, this data are in Section~\ref{sec:discussion}.

\begin{table*}
\centering
\caption{Summary of Laboratory Experience Feedback for Group MM}
\label{tbl:rawLikertData}
\begin{tabular}{|c|c|c|c|c|c|c|c|}
\hline
             & \textbf{Strongly Agree} & \textbf{Agree} & \textbf{Somewhat Agree} & \textbf{Neither Agree nor Disagree} & \textbf{Somewhat Disagree} & \textbf{Disagree} & \textbf{Strongly Disagree} \\ \hline
\textbf{Q3}  & 76\%                    & 17\%           & 5\%                     & 0\%                                 & 0\%                        & 2\%               & 0\%                        \\ \hline
\rowcolor[HTML]{C0C0C0} 
\textbf{Q4}  & 0\%                     & 2\%            & 2\%                     & 2\%                                 & 5\%                        & 17\%              & 71\%                       \\ \hline
\textbf{Q5}  & 45\%                    & 24\%           & 7\%                     & 19\%                                & 2\%                        & 2\%               & 0\%                        \\ \hline
\rowcolor[HTML]{C0C0C0} 
\textbf{Q6}  & 0\%                     & 0\%            & 0\%                     & 2\%                                 & 2\%                        & 26\%              & 69\%                       \\ \hline
\rowcolor[HTML]{C0C0C0} 
\textbf{Q7}  & 2\%                     & 0\%            & 7\%                     & 10\%                                & 7\%                        & 31\%              & 43\%                       \\ \hline

\textbf{Q8}  & 62\%                    & 26\%           & 7\%                     & 0\%                                 & 2\%                        & 2\%               & 0\%                        \\ \hline
\rowcolor[HTML]{C0C0C0} 
\textbf{Q9}  & 7\%                     & 0\%            & 0\%                     & 2\%                                 & 7\%                        & 36\%              & 48\%                       \\ \hline
\textbf{Q10} & 48\%                    & 24\%           & 17\%                    & 10\%                                & 0\%                        & 2\%               & 0\%                        \\ \hline
\end{tabular}
\end{table*}

\vspace{-3.5mm}
\subsection{Quantitative - Laboratory Performance}

This section contains the aggregate data for each of the 12 laboratory assignments. It further contrasts the performance of each assignment and the total score for both the SM and MM groups.  It presents and considers the median data, as there are often data outliers on both ends of the spectrum. Most importantly, when students elected not to complete a lab or miss it, they received a zero grade. 

\begin{figure}
    \centering
    \includegraphics[width=\columnwidth]{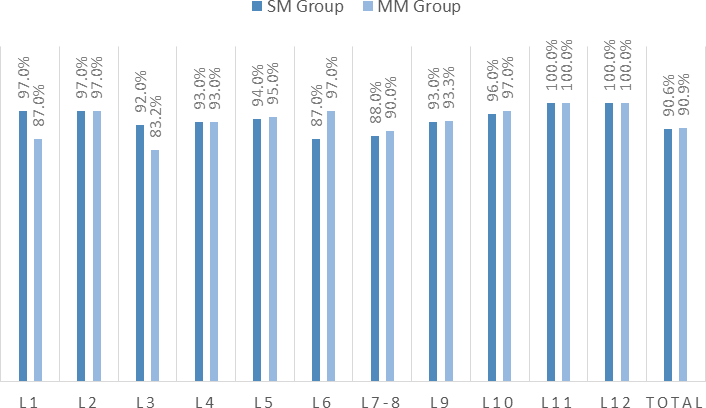}
    \caption{Median Laboratory Scores}
    \label{fig:medianLabScores}
\end{figure}

\begin{figure}
    \centering
    \includegraphics[width=\columnwidth]{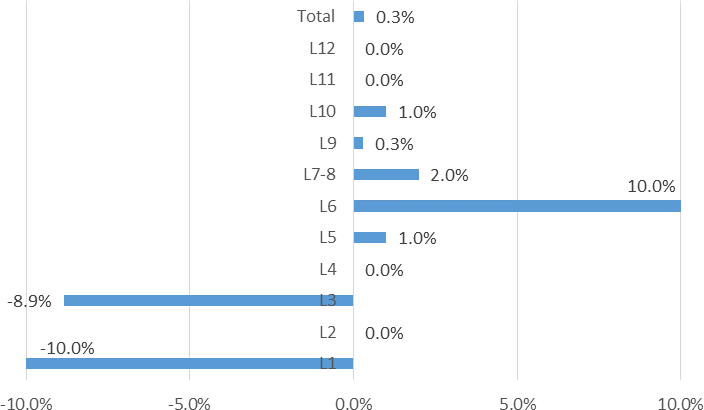}
    \caption{Difference between MM and SM Laboratory Scores}
    \label{fig:diffLabScores}
    \vspace{-5mm}
\end{figure}

The two groups are presented in Figure~\ref{fig:medianLabScores}. L\# represents the specific laboratory session.  The median grade for SM is in the darker shade, and MM in the lighter shade. Figure~\ref{fig:diffLabScores} illustrates the difference in scores for the MM group with respect to the SM group (MM-SM).   In 4 different laboratory assignments, there was an improvement of 1\% or more for the MM group, and an additional assignment with an improvement of 0.3\%. In 4 of the laboratory sessions, there was no difference in performance. There were 3 assignments in which the SM group performed better, Laboratories 1 and 3. Statistical analysis and insights into all this data can be found in Section~\ref{sec:discussion}, including the nature of the specific laboratory assignments that exhibited notable differences and statistical significance.  The total laboratory semester scores demonstrated an improvement of 0.3\% for the MM group.
\vspace{-3.5mm}
\subsection{Qualitative - Laboratory Experience}

This section summarizes the responses of the students when asked open-ended, qualitative, questions about their laboratory experiences. In order to avoid 'cherry picking' responses from the data, the study employed Heimerl et al.'s~\cite{heimerl2014word} of using word counting programs and wordcloud generators\footnote{\url{https://www.wordclouds.com/}} to perform text analytics.  This approach is demonstrably efficient at extracting and illustrating emerging and prevalent ideas and thoughts.  All of the students anonymized responses are available on a publicly available data set\footnote{\url{http://hdl.handle.net/2374.MIA/6612}}. Students were asked three specific questions and one completely open-ended question:

\begin{enumerate}
    \setcounter{enumi}{10}
    \item List out and describe any/all the ways in which you used the multiple monitors. What content did you have on each monitor?
    \item Describe ways in which you found multiple monitors assisted your learning.
    \item Which aspects of using multiple monitors did you find challenging?  Did they hinder your learning in any way?
    \item What other comments, observations, or conclusions do you have?
\end{enumerate}

A tabular representation of the students' prevalent responses/concepts is in tables~\ref{tbl:utilizationTable} to \ref{tbl:challenging}.  The corresponding word clouds and the interpretation are in the discussion in Section~\ref{sec:discussion}.

\begin{table}
\centering
\caption{Top 15 Words in Students' Utilization}
\label{tbl:utilizationTable}
\begin{tabular}{|c|c|}
\hline
\textbf{Word} & \textbf{Occurrences} \\ \hline
monitor       & 48                  \\ \hline
one           & 30                  \\ \hline
eclipse       & 25                  \\ \hline
lab           & 19                  \\ \hline
left          & 12                  \\ \hline
right         & 11                  \\ \hline
screen        & 11                  \\ \hline
assignment    & 10                  \\ \hline
code          & 10                  \\ \hline
requirements  & 10                  \\ \hline
open          & 9                   \\ \hline
lecture       & 8                   \\ \hline
put           & 8                   \\ \hline
instructions  & 7                   \\ \hline
notes         & 5                   \\ \hline
\end{tabular}
\vspace{-5mm}
\end{table}

\begin{table}
\centering
\caption{Top 20 Words in Students' Assistance}
\label{tbl:Assistance}
\begin{tabular}{|c|c|}
\hline
\textbf{Word} & \textbf{Occurrences} \\ \hline
one           & 14                  \\ \hline
able          & 13                  \\ \hline
code          & 12                  \\ \hline
easier        & 11                  \\ \hline
back          & 10                  \\ \hline
time          & 10                  \\ \hline
look          & 9                   \\ \hline
forth         & 8                   \\ \hline
screen        & 8                   \\ \hline
work          & 8                   \\ \hline
allowed       & 7                   \\ \hline
monitors      & 7                   \\ \hline
multiple      & 7                   \\ \hline
much          & 6                   \\ \hline
notes         & 6                   \\ \hline
switch        & 6                   \\ \hline
windows       & 6                   \\ \hline
also          & 5                   \\ \hline
easily        & 5                   \\ \hline
information   & 5                   \\ \hline
\end{tabular}
\end{table}

\begin{table}
\centering
\caption{Top 15 Words in Students' Challenges}
\label{tbl:challenging}
\begin{tabular}{|c|c|}
\hline
\textbf{Word} & \textbf{Occurrences} \\ \hline
monitors      & 10                   \\ \hline
challenging   & 8                    \\ \hline
hinder        & 8                    \\ \hline
learning      & 7                    \\ \hline
multiple      & 7                    \\ \hline
using         & 6                    \\ \hline
find          & 5                    \\ \hline
None          & 5                    \\ \hline
screen        & 5                    \\ \hline
think         & 5                    \\ \hline
window        & 4                    \\ \hline
nothing       & 3                    \\ \hline
one           & 3                    \\ \hline
open          & 3                    \\ \hline
time          & 3                    \\ \hline
\end{tabular}
\vspace{-5mm}
\end{table}

\vspace{-4mm}
\section{Discussion}
\label{sec:discussion}

%The purpose of the discussion is to interpret and describe the significance of your findings in light of what was already known about the research problem being investigated, and to explain any new understanding or fresh insights about the problem after you've taken the findings into consideration. The discussion will always connect to the introduction by way of the research questions or hypotheses you posed and the literature you reviewed, but it does not simply repeat or rearrange the introduction; the discussion should always explain how your study has moved the reader's understanding of the research problem forward from where you left them at the end of the introduction.

\subsection{Quantitative - Laboratory Experience via Likert-type Scale}

In answering research question one, the two figures, Figure~\ref{fig:positiveLikert} and Figure~\ref{fig:negativeLikert},  illustrate a very supportive and positive overall laboratory experience for the students.  Some very minor exceptions include Question 5 and 7. Question 5 (positively phrased question) in Figure~\ref{fig:positiveLikert} pertaining to agreeing that "programming should be taught by having students use multiple monitors" appears slightly lighter than the rest. Question 7 (negatively phrased question), as seen in Figure~\ref{fig:negativeLikert}, referring to deciding which content goes to which monitor, appears slightly darker than the others. For the positive (3,5,8, and 10) questions, there is strong agreement in the range of 45-76\%, and for the negative (4,6,7, and 9) questions, there is strong disagreement in the range of 43-71\%. Notably, 20\% of students neither agree nor disagree that programming should be taught to students using multiple monitors.  To assist with readability, the "negatively phrased" questions are shaded in Table~\ref{tbl:rawLikertData}. Thus, in the shaded rows, disagreement is more supportive of multiple monitors. In the non-shaded rows, agreement is more supportive of multiple monitors. 

Despite the lack of strong supporting performance data, as discussed in the next subsection, it is evident from both Figures \ref{fig:positiveLikert} and \ref{fig:negativeLikert} that students found their laboratory experience was enhanced and improved by the implementation of multiple monitors at their work stations.  Most notably, from the positively phrased questions, there was strong agreement in questions 3 and 8. For question 3, 76\% of the students strongly agree that they would chose to use multiple monitors for programming again in the future, with only 7\% choosing the option somewhat agree or lower, and 2\% choosing an option of neutral or lower. For question 8, 62\% strongly agree that they found using multiple monitors to be intuitive and easy to learn, with only 11\% choosing the option somewhat agree or lower, and 4\% choosing an option of neutral or lower.  Question 5 also illustrated very strong support for multiple monitors, with only 9\% of respondents choosing an option of somewhat agree or lower that programming should be taught by having students employ multiple monitors. 
\vspace{-2.5mm}
\subsection{Quantitative - Laboratory Performance}

This subsection begins with data analysis and follows that with an interpretation of the data to address the second research question.

\subsubsection{Data Analysis}

With the assistance of Miami University's data analysis group, Research Computing Support (RCS) \footnote{\url{https://www.miamioh.edu/research/research-computing-support/}}, the following is statistical analysis on the laboratory performance data using Mood’s Median Non Parametric Hypothesis Test~\cite{corder2014nonparametric}.  This includes performing tests on the medians instead of the means, since there  were a significant number of low scores on the different lab assignments and that would violate the normality assumption required in conducting chi-squared test on means~\cite{corder2014nonparametric}.

Overall, comparing the medians of labs SM with MM, the medians for only Labs 1 thru 3 were higher for sections SM. Starting with Lab 4, and all the way to the final Lab, 12, the medians were equal or higher for sections MM. As illustrated in Table~\ref{tbl:p-values}, the median analysis yielded only 3 labs demonstrating statistical significance based on having chi-squared test P-Values
\textless 0.05: Labs 1, 3, and 6. Thus, only those data graphs from only those labs' are presented here. However, the entire detailed data analysis, including P-Value calculation for all labs, can be found on the public data set. The next subsection contains a discussion on the labs in general.  

\begin{table}
\centering
\caption{Summary of P-Values}
\label{tbl:p-values}
\begin{tabular}{|c|c|}
\hline
\textbf{Lab \#} & \textbf{Pr\textgreater Chi-Square} \\ \hline
1               & 0.0003                              \\ \hline
2               & 0.3899                              \\ \hline
3               & 0.0016                              \\ \hline
4               & 0.8090                              \\ \hline
5               & 0.4553                              \\ \hline
6               & \textless{}0.0001                   \\ \hline
7-8             & 0.6673                              \\ \hline
9               & 0.6611                              \\ \hline
10              & 0.3815                              \\ \hline
11              & 0.1662                              \\ \hline
12              & 0.1740                              \\ \hline
\end{tabular}
\vspace{-3mm}
\end{table}

\begin{figure}
    \centering
    \includegraphics[width=\columnwidth]{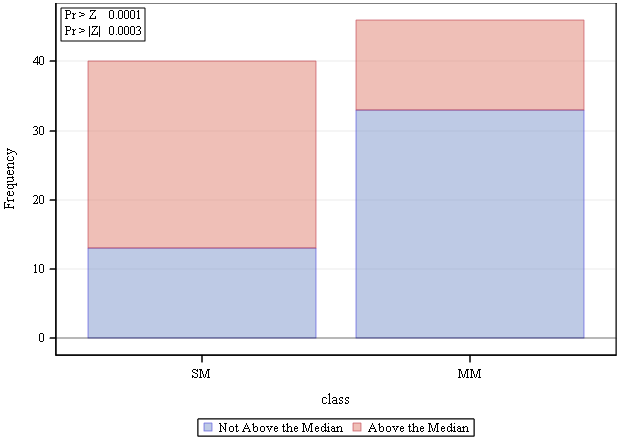}
    \caption{Lab 1 Frequencies Above \& Below Overall Median}
    \label{fig:lab1}
    \vspace{-5mm}
\end{figure}

\begin{figure}
    \centering
    \includegraphics[width=\columnwidth]{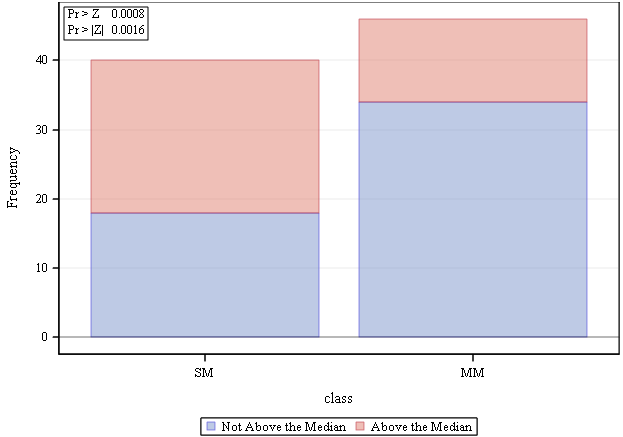}
    \caption{Lab 3 Frequencies Above \& Below Overall Median}
    \label{fig:lab3}
    \vspace{-5mm}
\end{figure}

\begin{figure}
    \centering
    \includegraphics[width=\columnwidth]{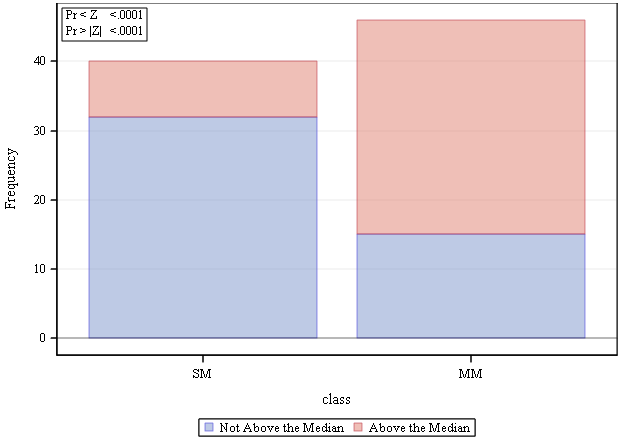}
    \caption{Lab 6 Frequencies Above \& Below Overall Median}
    \label{fig:lab6}
    \vspace{-5mm}
\end{figure}

Labs 1 and 3, which are showcased in Figures~\ref{fig:lab1} and \ref{fig:lab3}, both illustrate statistical significance in improved student performance using single monitor work stations. The pink (lighter) shade represents scores above the median, while the blue represents those below the median.  Lab 6 showcased the most significant result of the entire empirical study, with a very low P-Value and very substantial performance improvement by the MM group as presented in Figure~\ref{fig:lab6}.

\subsubsection{Data Interpretation}
Overall, there was a slight total performance improvement in the MM scores. 4 of the individual laboratories had notable improvement, however, only 1 of those 4 exhibited statistical significance.  The 4 equal and 2 negative performance differences in the MM group weaken any strong correlation.  One potential explanation for the difference in the SM group outperforming MM in Laboratories 1 to 3 is that the MM group, of which roughly 66\%  had not programmed using multiple monitors before, were acclimating to them during a timed laboratory session. However, 1) one would expect a continual improvement, which the data does not support, and 2) the students' subjective experiences appear to indicate they found it easy and intuitive, as demonstrated in Question 8.  However, there may be confirmation and recognition bias in their responses. An interesting consideration is the types of assignments.  For example, Laboratories 5-8 were software design related, and 9 and 10 were graphics related.  Perhaps these types of activities are more conducive to multiple monitor environments, which is an interesting avenue of tailored and granular future research. 

There were a number of statistically significant labs. Laboratory 1, on which the SM group performed better, was entirely a review of the prerequisite course.  It is possible that the first time use of multiple monitor work stations for the 66\% of first-timers added complexity and distracted the students from performing well on the review. That is, the SM group had to focus only recall in their brains while the MM group had to focus on both recall and new learning/experiences as they adjusted to multiple monitors. Laboratory 2, while not statistically significant, was also a "review" lab that demonstrated better SM performance. Laboratory 3 had the SM group performing better than the MM group but less significantly so than Laboratory 1. It focused on the basics of object oriented programming by implementing 3 basic classes and their variables and simple methods.  This was the students' first time using object orientation in practice. One would suspect that having the assignment requirements and their class definitions on one monitor and the code on the other monitor would improve performance, but the data does not support this. The remaining labs, 4 to 12, exhibited better performance for the MM group. Laboratory 6, which involved software design using interface definition with multiple classes and hierarchies, was the most statistically significant finding in the study.   This is as expected because the assignment involves many "nouns" that are class candidates to form interfaces and class hierarchies. Multiple context switches would be required by students in a single monitor work station as the brain must parse the text from the assignment requirements and then must codify these relationships and interfaces using the programming language. 

The fact that the majority of students have not programmed using multiple monitors prior to this intervention is noteworthy in itself. It also provides context for the relatively less favourable, darker in Figure~\ref{fig:negativeLikert}, response to Question 7 pertaining to the students being unsure of which content to put on which monitor.  The study made a conscious decision to not have a tutorial on using multiple monitors nor provide formal suggestions because 1) of the full course schedule, 2) a desire to observe how the students would use them intuitively, as would be the situation in most educational intuitions. 

Another interesting observation pertains to Question 5.  Roughly a fifth of the students "neither agree nor disagree" that programming should be taught to students using multiple laboratory monitors despite over three quarters of them saying they would choose to program using multiple monitors.  Possible explanations for any intersections may include the students not wanting to speak for others, or they consider general programming and educational programming/learning to be different enough to not warrant multiple monitors in both cases. 
\vspace{-2.5mm}
\subsection{Qualitative - Laboratory Experience}
Using Heimerl et al.'s~\cite{heimerl2014word} approach of employing word clouds to perform text analytics, the third research question is explored and as are emergent concepts and ideas held by the students. The word clouds were generated through a publicly available web tool\footnote{\url{https://www.wordclouds.com/}}.  Herein, this paper quotes words that appear directly and prominently in the corresponding laboratory experience student responses. For rigor and verification, all of the anonymized student responses can be found in the public data set. 

\begin{figure}
    \centering
    \includegraphics[width=\columnwidth]{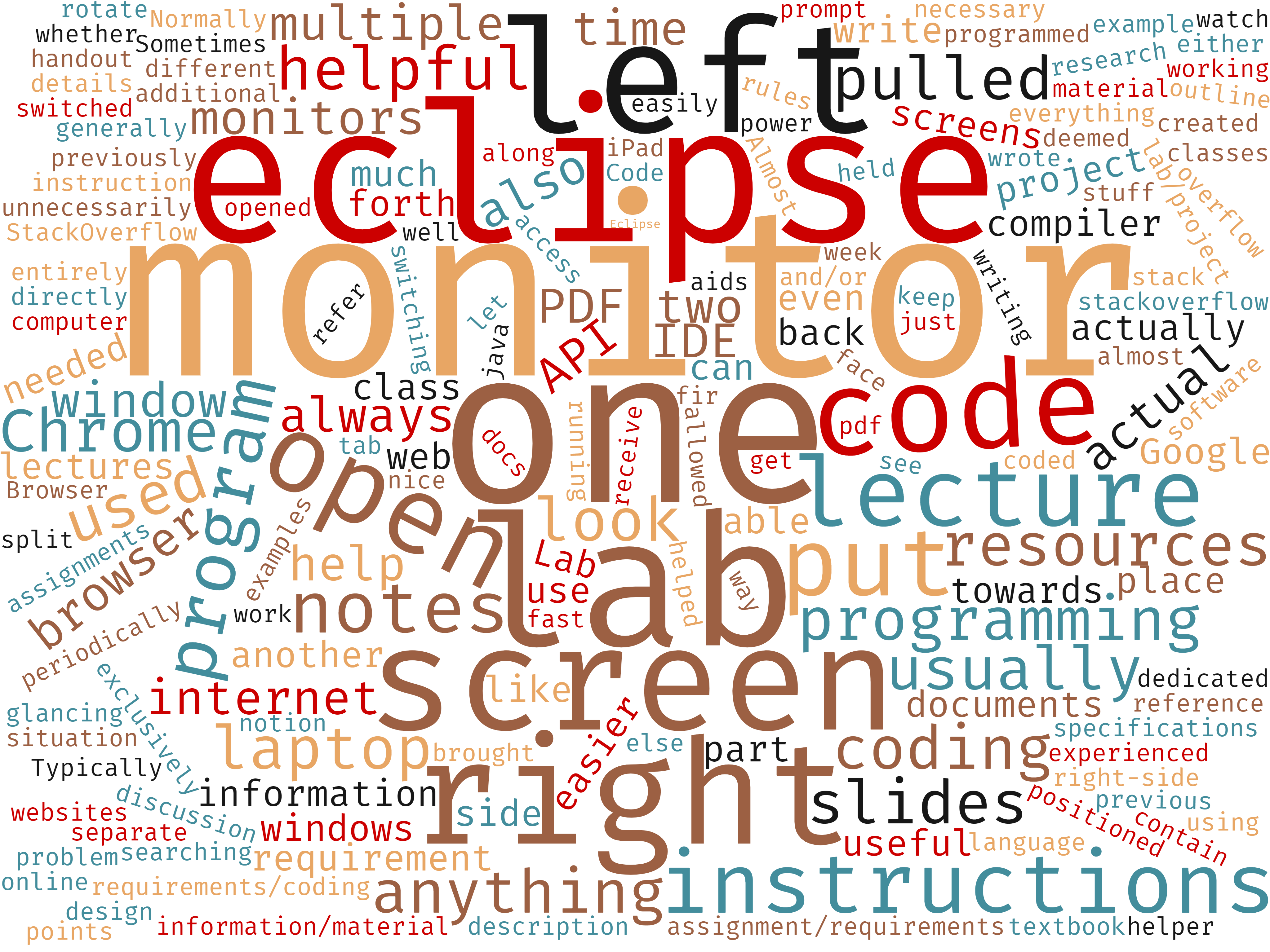}
    \caption{Response Words about Content and Use}
    \label{fig:contentResponses}
        \vspace{-2.5mm}
\end{figure}

\begin{figure}
    \centering
    \includegraphics[width=\columnwidth]{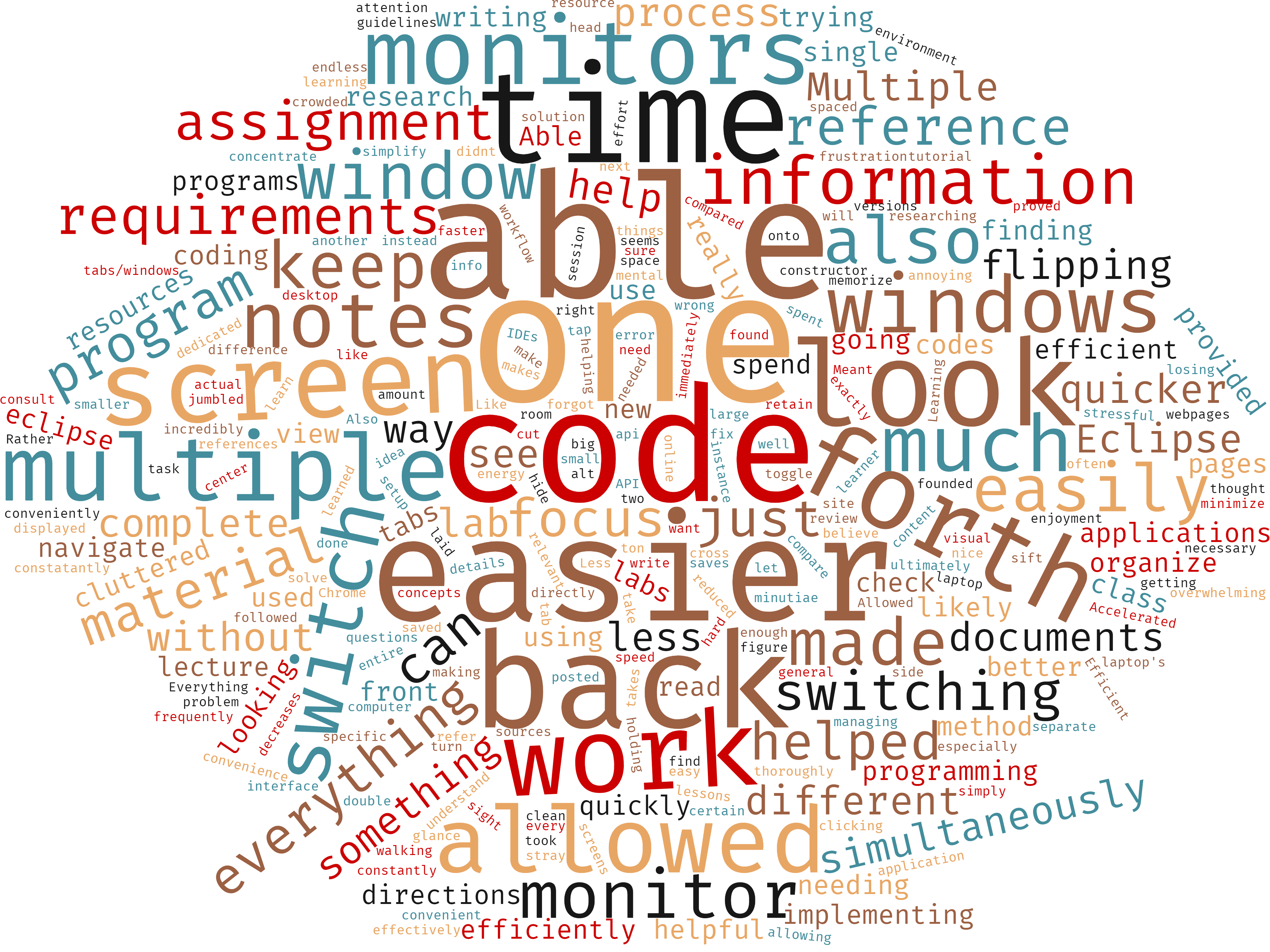}
    \caption{Response Words about Learning Assistance}
    \label{fig:learningCloud}
        \vspace{-5mm}
\end{figure}

\begin{figure}
    \centering
    \includegraphics[width=\columnwidth]{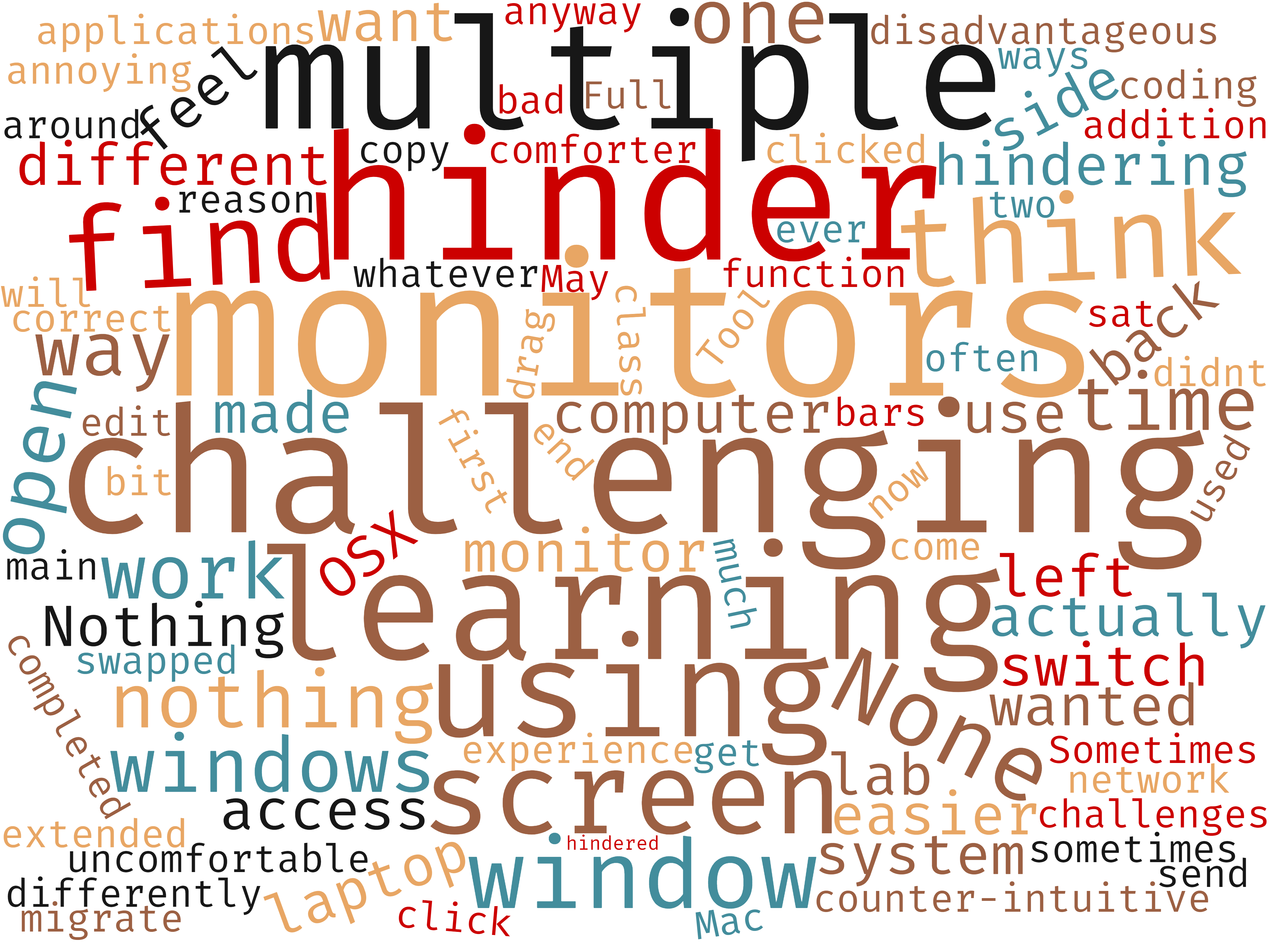}
    \caption{Response Words about Challenges \& Hindrances}
    \label{fig:hinderCloud}
    \vspace{-5mm}
\end{figure}

The students' answer to the content and the ways in which they used the multiple monitor setup is illustrated in Figure~\ref{fig:contentResponses}. Students discussed putting some combination of their "Eclipse" development environment and "code", the "assignment" and its "requirements" / "instructions", and their "lecture" "notes" on the "left" and "right" "monitors". For the ways in which the students perceived it to assist their learning, a representation of their responses is in Figure~\ref{fig:learningCloud}. Many students cited their tasks being "easier", them saving "time", and being more "able" / "allowed" to perform and complete tasks.  There is also much mention of what they "did not" have to do, including not having to "switch(ing)" "back" and "forth".  Question 13 about how using multiple monitors hindered their learning or aspects they found challenging yielded responses mostly denying any such negative experiences. As visualized in Figure~\ref{fig:hinderCloud}, much of them mentioned how they found "nothing" / "none" of it "challenging", and that it "did not" "hinder" their laboratory experience nor learning. Question 14 is omitted as they were too varied to have any consistently emerging ideas and on different topics, but is included in the public data set for reader review.

    \vspace{-2.5mm}
\subsection{Considerations and Threats to Data Validity}

Best efforts were made to have consistency for all students, except for the number of monitors.  The instructor was the same for all students, the teaching assistants rotated for an equal distribution of laboratory supervision and grading, and all students had the same amount of time to do the labs. However, there are potential threats to the empirical study's data. 

The main threat is the organically occurring distribution and makeup of the students in the control group versus the intervention group.  The author had no influence as to which students enrolled in which section, and the students had no prior knowledge about potential interventions before the semester began.  This distribution introduces a number of threats and considerations to the data. The first aspect relates to gender, as it was not considered nor accounted for gender in the data.  Gender impacts computer supported learning~\cite{gunn2003dominant}, and different genders can learn better using different learning styles~\cite{lau2009exploring}.  Thus, it is possible that multiple monitors could have been used and appreciated differently by different genders. Another aspect threatening the data is the perquisite grade each student had entering the course.  Having the perquisite grade criterion ensured that all student subjects had the same \textbf{minimum} knowledge level. However, there is potentially high variance in their previous course performance above that minimum required grade and other programming activities and experiences, such as programming clubs and side projects. Lastly, there were six more students in the MM group.  Overall, since there was no intention to influence nor interfere with enrollment out of fairness to the students, and this is merely an empirical study, these are acceptable threats. Potential future work includes considering and isolating factors like these and others and their respective impacts on using multiple monitors in educational laboratory work stations.

\vspace{-2mm}
\section{Conclusion}
\label{sec:conlusion}

While more is not always better, it is seemingly intuitive that an additional monitor and more digital space would improve computing students' laboratory experience and, potentially, their performance.  It also better prepares students for the work environments they will encounter in the real world. However, due to myriad factors, not all students have access to work stations with multiple monitors nor can educational institutions provide them.  Through a technology grant, Miami University was able to facilitate multiple monitors to the main laboratory only. This provided an organic control group, as scheduling and other constraints meant that some sections of the course were limited to single-monitor work stations.  This paper presented an empirical study in the context of an introductory object-oriented course, featuring two sections with the multiple monitor intervention, and two with single monitor work stations.   Only a slight total performance improvement was observed for the intervention group, but there were more notable improvements in the laboratory assignments dealing with graphics and software design.  With statistical significance, it can be concluded that the students using the single monitor work stations performed better on review and introduction lab assignments. However, the most statistically significant result was the students using multiple monitor work stations performed much better on a lab featuring interface definitions and class hierarchies.  The students from the intervention group rated their laboratory experience with the intervention very favourably, as demonstrated in both their quantitative and qualitative responses. Thus, while the performance data does not necessarily nor strongly support the use of multiple monitors, the students' experiences were extremely enhanced. One can argue that enjoyment and gratification is just as important as performance regarding a student's university experience.  It is the intent of this empirical study to help support the case for providing multiple monitor work stations for all students, and to guide future research in employing multiple monitors as an instructional technology for early programming education. 
    \vspace{-2.5mm}
% or


%\section*{Acknowledgment}
%We firstly and foremost want to thank the Research Computing Support (RCS) at Miami University for all their help and assistance with our data analysis, and also working with us to generate our data graphs. We specifically want to thank Jon Patton from that group for their help with our study analysis. The multiple monitor setup would not have been possible without the financial support of Miami University and its Information Technology Department through its Student Technology Fee Grant. We would also like to thank our institution's Institutional Review Board for helping ensure we comply with federal regulations and our institution policies on research involving human subjects and ethics. Lastly, we want to extend our thanks to all students who volunteered to include their grades in our performance data, and those in the MM group that chose to complete the optional survey. 


\ifCLASSOPTIONcaptionsoff
  \newpage
\fi



%\bibliographystyle{IEEEtran}

\bibliographystyle{IEEEtran}
\bibliography{bare_jrnl}


%\end{IEEEbiography}


%\newpage



%\vfill


\end{document}